\documentclass[12pt]{article}
\usepackage{amssymb}
\usepackage{amsmath}
\usepackage[space]{cite}
\usepackage{algorithm}
\usepackage{bbm}
\usepackage[noend]{algpseudocode}
\usepackage[normalem]{ulem}
\usepackage{url}
\usepackage{graphicx}
\usepackage{adjustbox}
\usepackage{caption}
\usepackage{subcaption}
\usepackage{enumitem}
\usepackage{xcolor}

\usepackage{MnSymbol}

\usepackage{cases}

\numberwithin{equation}{section}

\newcommand{\non}{\nonumber}
\newcommand{\id}{\mathbb{I}}

\newcommand{\cut}[1]{\ifmmode\text{\textcolor{red}{\sout{\ensuremath{#1}}}}\else\textcolor{red}{\sout{#1}}\fi}

\newcommand{\RN}[1]{%
  \textup{\uppercase\expandafter{\romannumeral#1}}%
}

\usepackage{quantikz}
\usetikzlibrary{circuits.logic.US}
\tikzset{
    gateO/.style={
        draw,
        circle,
        minimum width=0.5em,
        inner sep=2pt    }
}
\DeclareExpandableDocumentCommand{\gateO}{O{}{m}}{|[gateO,#1]| {#2} \qw}

\tikzset{
    gateOS/.style={
        draw,
        circle,
        minimum width=0.5em,
        inner sep=2pt,
		fill=red!20}
}
\DeclareExpandableDocumentCommand{\gateOS}{O{}{m}}{|[gateOS,#1]| {#2} \qw}

\begin{document}

\begin{titlepage}
\vspace{.5in}
\begin{center}

{\LARGE Dicke states as matrix product states}\\
\vspace{1in}

\large David Raveh\footnote{Department of Physics and Astronomy, Rutgers University, Piscataway, NJ 08854-8019 USA, {\tt 
david.raveh@rutgers.edu}} and 
Rafael I. Nepomechie\footnote{
Department of Physics, P.O. Box 248046, University of Miami,
Coral Gables, FL 33124 USA, {\tt nepomechie@miami.edu}}

\end{center}

\vspace{.5in}

\begin{abstract}
We derive an exact canonical matrix product state (MPS) representation for Dicke states $|D^n_k\rangle$ with minimal bond dimension $\chi=k+1$, for general values of $n$ and $k$, for which the W-state is the simplest case $k=1$.
We use this MPS to formulate a quantum circuit for sequentially preparing Dicke states deterministically, relating it to the recursive algorithm of B{\"a}rtschi and Eidenbenz. We also find exact canonical MPS representations with minimal bond dimension for higher-spin and qudit Dicke states.
\end{abstract}

\end{titlepage}

\setcounter{footnote}{0}

\section{Introduction}\label{sec:intro}

Matrix product state (MPS) representations of states of one-dimensional quantum many-body systems have important applications, both conceptually (such as capturing the structure of entanglement, and establishing the physical basis of the density matrix renormalization group) and practically (such as performing explicit computations, and preparing states sequentially on a quantum computer), see e.g. \cite{Vidal:2003lvx, Schon:2005, Perez-Garcia:2006nqo, Schollwock:2011} and references therein. MPS representations are typically obtained numerically. However, the exact MPS representation for the AKLT state \cite{Affleck:1987vf} has had a particularly profound impact. Indeed, it has made possible the exact computation of correlation functions in the AKLT state, as well as the preparation of this state on quantum computers. One recent example of the latter is the preparation of AKLT states in constant depth on a quantum computer using intermediate measurement and feed-forward techniques \cite{Smith:2023}. (Quantum state preparation based on a numerical MPS with large bond dimension appears challenging since the needed 2-site unitaries and their gate decompositions must be determined individually for each site.) Unfortunately, exact MPS expressions are known for only a handful of other states, such as GHZ and cluster states, see e.g. \cite{Perez-Garcia:2006nqo}; and these examples all have low bond dimension $\chi=2$. It could be valuable to identify more examples of exact MPS representations, especially for states with higher bond dimension. Indeed, such examples could be used for computing correlation functions, and for quantum state preparation; moreover, they could serve as simple toy models for treating more complicated states. (As is well known, MPS-inspired quantum state preparation methods, with a given bond dimension, can avoid the problem of exponential scaling of resources with system size.)

In this article, we derive an exact canonical MPS representation for  Dicke states $|D^n_k\rangle$, which has minimal bond dimension $\chi=k+1$, for general values of $n$ and $k$. These states have been extensively studied and exploited in quantum information and computation for numerous tasks, including quantum networking, quantum metrology, quantum tomography, quantum compression, and optimization, see e.g. \cite{Dicke:1954zz, Murao:1999, Childs:2000, Popkov:2004, Latorre:2004qn, Ozdemir:2007, Prevedel:2009, Toth:2010, Toth:2012, Lamata:2013, Farhi:2014, Ouyang:2014, Ouyang:2021, Bartschi2019, Wang:2021, Buhrman:2023rft, Bond:2023zuy, Piroli:2024ckr}. Features of Dicke states that make them particularly useful include their robustness against decoherence \cite{Guhne2008}, their permutation symmetry that facilitates tomography \cite{Toth:2010}, their entanglement, and their relative simplicity.
We also find exact MPS representations for higher-spin Dicke states \cite{Liu:2015, Nepomechie:2024fhm}, as well as for qudit Dicke states \cite{Wei:2003, Popkov:2005, Hayashi:2008,  Wei:2008, Zhu:2010, Carrasco:2015sxh, Aloy:2021,Li:2021, Nepomechie:2023lge}. Our main tool for all the cases (qubit, higher-spin, and qudit Dicke states)
is the Schmidt decomposition, which for Dicke states is particularly simple, as it is dictated by the Clebsch-Gordan theorem. The resulting MPS representations for the various cases are all qualitatively similar. An interesting feature of the higher-spin Dicke MPS is that the bond dimension is 
given by $\chi=k+1$, independently of the value of the spin; while the bond dimension of the qudit Dicke MPS depends more intricately on the occupation numbers $\vec k$, see Table \ref{table:dims}.

We use the qubit MPS to formulate a quantum circuit for sequentially preparing qubit Dicke states deterministically,
with $\mathcal{O}(k\,n)$ size and depth. This circuit closely corresponds to the recursive algorithm given by B{\"a}rtschi and Eidenbenz \cite{Bartschi2019}, which has similar size and depth. Indeed, our circuit can be regarded as an MPS realization of the latter. However, the latter circuit has the advantage of not requiring an ancilla qudit, and is therefore more practical. (Shallower, but probabilistic, circuits for preparing these Dicke states are known \cite{Wang:2021, Buhrman:2023rft, Piroli:2024ckr}.)

We note that the MPS representations presented here are not strictly canonical, but are `sufficiently' so, in the sense that they can be used for sequential state preparation. These MPSs can be made strictly canonical by adding suitable correction terms.

There has been surprisingly little earlier work on exact MPS formulations of Dicke states, at least to our knowledge. Indeed, an exact MPS is known for W-states  \cite{Perez-Garcia:2006nqo}, which are the simplest Dicke states $|D^n_k\rangle$ with $k=1$; however, that MPS is not canonical, see also \cite{Klimov:2023srk}.
Building on \cite{Sanz:2016}, an MPS for a linear combination of Dicke states is studied in \cite{Aloy:2021}. However, that construction requires solving a system of equations, whose solution appears to be singular in the limit of a single Dicke state; moreover, that MPS is not canonical, and has a higher bond dimension than the MPS presented here. The works \cite{Sanz:2016, Aloy:2021} restrict to MPS representations that are translationally invariant; we do not impose this requirement on our MPS, which allows us to achieve the results presented here.

The remainder of the paper is organized as follows. In Sec. 
\ref{sec:general}, we briefly review some key facts about MPS representations. In Sec. \ref{sec:su2}, we first obtain an exact canonical MPS for ordinary (qubit, or spin-1/2) Dicke states; and we then generalize this construction to Dicke states of arbitrary half-integer spin $s=1/2, 1, 3/2, \ldots$. In Sec. \ref{sec:sud}, we obtain an analogous MPS for qudit Dicke states. We briefly discuss these results in Sec. \ref{sec:discuss}. In Appendix \ref{sec:circuit} we 
formulate a quantum circuit for sequentially preparing Dicke states deterministically based on the MPS obtained in Sec. \ref{sec:su2}. Code in cirq \cite{cirq} for simulating this circuit is provided in the Supplemental Material \cite{SuppMat}.

\section{Generalities}\label{sec:general}

Let us consider a system of $n$ qudits, where each qudit is multi-leveled, with $d$ levels. The Hilbert space of this system is therefore $\mathcal{H}=(\mathbb{C}^{d})^{\otimes n}$. In the computational basis, a state $|\psi\rangle \in \mathcal{H}$ is expressed as a sum over $d^n$ basis states with complex coefficients:
\begin{equation}
|\psi\rangle=\sum_{m_1,\dots ,m_n=0}^{d-1} a_{m_1\dots m_n}|m_n\dots m_2 m_1\rangle\equiv \sum_{\vec m} a_{\vec m}\,|\vec m\rangle.
\end{equation}
We consider representing the coefficients in the form $a_{\vec m}=\langle \underline L|A_n^{m_n}\dots A_2^{m_2}A_1^{m_1}|\underline R\rangle$, so that 
\begin{equation}
|\psi\rangle=\sum_{\vec m}\langle\underline L|A_n^{m_n}\dots A_2^{m_2}A_1^{m_1}|\underline R\rangle\,|\vec m\rangle\,,
\label{genMPS}
\end{equation}
where $A_i^{m_i}$ are $\chi\times\chi$ matrices. Such a state is called a \emph{matrix product state} (MPS) of \emph{bond dimension} $\chi$ with \emph{open boundary conditions}, determined by the boundaries $|\underline L\rangle$ and $|\underline R\rangle$. Here and in what follows, we underline a vector to emphasize that it is a $\chi$-dimensional ancillary vector, as opposed to the $d$-dimensional vectors (not underlined) corresponding to system qudits. In the special case that $A_i^m\equiv A^m$, i.e. the MPS is site-independent, the MPS is referred to as \emph{translationally invariant} in the bulk (neglecting the boundaries $|\underline L\rangle,|\underline R\rangle$).

Every state $|\psi\rangle$ can be expressed as an MPS given large enough $\chi$, although non-uniquely \cite{Schollwock:2011}; the smallest such $\chi$ is given by the Schmidt rank of the state $|\psi\rangle$, i.e. the maximum number of terms in the Schmidt decompositions of $|\psi\rangle$ (maximized over all possible cuts) \cite{Dalzell:2019}. Further, one can always find an MPS in \emph{left-canonical form}, i.e. for each site $i$ we have \cite{Vidal:2003lvx, Perez-Garcia:2006nqo, Schollwock:2011}
\begin{equation}
    \sum_m {A_i^m}^\dag A_i^m = \id\,;
\end{equation}
we shall simply refer to such an MPS as being `canonical'. This condition implies that there exist two-qudit unitary operators $U_i$ acting on an ancillary qudit $|\underline j\rangle$ and the system qudit at site $i$, performing the mapping
\begin{equation}
U_i\,|\underline j\rangle\,|0\rangle_i=\sum_m (A_i^m\,|\underline j\rangle)\,|m\rangle_i
\label{Ugen}
\end{equation}
for all $|\underline j\rangle$. This gives a natural method of preparing the state $|\psi\rangle$ on a quantum computer with sequential unitary operations: 
beginning with $n$ system qudits ($d$-level) and one ancillary qudit of dimension $\chi$, applying the two-qudit operators $U_n\dots U_2U_1 $ on the product state $|\underline R\rangle \, |0\rangle^{\otimes n}$ gives
\begin{equation}
U_n\dots U_2U_1\,|\underline R\rangle\,|0\rangle^{\otimes n}=\sum_{\vec m}(A_n^{m_n}\dots A_2^{m_2}A_1^{m_1}|\underline R\rangle)\,|\vec m\rangle=
|\underline L\rangle\, |\psi\rangle\,,
\label{genprep}
\end{equation}
where we have chosen our $A^m_i$ matrices so that the ancilla qudit decouples from the system qudits after applying our $n$ unitaries (this can always be done, see e.g. \cite{Schon:2005}).

\section{$SU(2)$ Dicke states}\label{sec:su2}

An ordinary qubit ($d=2$) Dicke state is a uniform superposition of all qubit computational basis states with a fixed number of 1's. Such a states can be expressed as 
\begin{equation}
    |D^n_k\rangle=\frac{1}{\sqrt{\binom{n}{k}}}\sum_{w\in P(n,k)} |w\rangle\,,
\label{Dickedef}
\end{equation}
where we sum over all permutations $w$ with $k$ ones and $n-k$ zeros. For simplicity, we restrict ourselves to $k\leq n/2$; similar analysis can be done for $k>n/2$. For example, with $(n,k)=(4,2)$ we have
\begin{equation}
    |D^4_2\rangle = \frac{1}{\sqrt{6}}\left(|00 1 1 \rangle + |0 1 0 1  \rangle +|01 1 0 \rangle+
    |100 1 \rangle + |1010  \rangle +|1 1 00 \rangle
    \right).
\end{equation}

We claim that a canonical MPS for the Dicke state $|D^n_k\rangle$ with minimal bond dimension $\chi=k+1$ is given by 
\begin{equation}
|D^n_k\rangle=\sum_{\vec m}\langle \underline{k}|A_n^{m_n}\dots A_2^{m_2}A_1^{m_1}|\underline 0\rangle\,|\vec m\rangle\,
\label{DickeMPS}
\end{equation} 
where $A_i^m$ are $(k+1) \times (k+1)$ quasi-diagonal matrices with elements 
\begin{equation}
\langle\underline j'|A_i^{m}|\underline j\rangle=\frac{c^{nk}_{i,j+m}c^{i,j+m}_{i-1,j}}{c^{nk}_{i-1,j}}\,\delta_{j',j+m}\,, \qquad 
 c^{nk}_{ij}=\sqrt{\frac{\binom{i}{j}\binom{n-i}{k-j}}{\binom{n}{k}}}\,,
\label{Dickeelem}
\end{equation}
where $c^{nk}_{ij}$ denotes the hypergeometric distribution.

This MPS can be derived as follows. We begin by writing the Schmidt decomposition for the Dicke states \cite{Latorre:2004qn}
\begin{equation}
|D^n_k\rangle=\sum_{j=\max(0,k-n+i)}^{\min(k,i)} c^{nk}_{ij}\,|D^{n-i}_{k-j}\rangle\,|D^i_j\rangle\,,
\label{DickeSchmidt}
\end{equation}
where we have partitioned our qubits into subsystems of sizes $n-i$ and $i$. The coefficients $c^{nk}_{ij}$ can be understood as $SU(2)$ Clebsch-Gordan coefficients.

The idea is to search for operators $U_i$ such that 
\begin{align}
U_i\dots U_2U_1\,|\underline 0\rangle\,|0\rangle^{\otimes n}&=\sum_j c^{nk}_{ij}\,|\underline j\rangle\,|0\rangle^{\otimes(n-i)}\,|D^i_j\rangle  \label{step1}\\
&=\sum_j c^{nk}_{ij}\,\sum_m c^{ij}_{i-1,j-m}\,|\underline j\rangle\,|0\rangle^{\otimes(n-i)}\,|m\rangle_i\,|D^{i-1}_{j-m}\rangle\,, \label{step1a}
\end{align}
where we pass to the second equality by performing the Schmidt decomposition on $|D^i_j\rangle$, and noting that $|D^1_m\rangle = |m\rangle$. In other words, the $U$ operators couple each term in the Schmidt decomposition to a different level in the ancilla $|\underline j\rangle$. The maximum number of terms in the Schmidt decomposition occurs at $i=\lfloor n/2\rfloor$, where there are $k+1$ terms; thus, for the Dicke states, the ancillary qudit requires dimension $\chi=k+1$.\footnote{The Schmidt rank of the Dicke states is invariant under $k\to n-k$; in general, $\chi=\min(k,n-k)+1$.}

Recalling \eqref{Ugen}, we next consider the ansatz \begin{equation}
U_i\,|\underline j\rangle\,|0\rangle_i =\sum_m (A_i^m\,|\underline j\rangle)\,|m\rangle_i 
= \sum_m\gamma^{(i)}_{j,m}\,|\underline{j+m}\rangle\,|m\rangle_i\,.
\label{ansatz}
\end{equation}
The reason for choosing this ansatz will soon be clear. It then follows that 
\begin{align}
\label{step2}
U_i U_{i-1}\dots U_2U_1\,|\underline 0\rangle\,|0\rangle^{\otimes n}&=U_i\sum_{j'} c^{nk}_{i-1,j'}\,|\underline {j'}\rangle\,|0\rangle^{\otimes(n-i+1)}\,|D^{i-1}_{j'}\rangle\\
&=\sum_{j'} c^{nk}_{i-1,j'}\sum_m \gamma^{(i)}_{j',m}\,|\underline{j'+m}\rangle\,|0\rangle^{\otimes(n-i)}\,|m\rangle_i\,|D^{i-1}_{j'}\rangle\,,\non
\end{align}
where the first equality follows from \eqref{step1} with $i \mapsto i-1$.
Equating \eqref{step1a} and \eqref{step2}, and enforcing $j'=j-m$, it follows that
\begin{equation}
c^{nk}_{ij}c^{ij}_{i-1,j-m}=c^{nk}_{i-1,j-m}\gamma^{(i)}_{j-m,m}\,
\end{equation}
so
\begin{equation}
\gamma^{(i)}_{j,m} =\frac{c^{nk}_{i,j+m}c^{i,j+m}_{i-1,j}}{c^{nk}_{i-1,j}} 
= \sqrt{\frac{\binom{n-i}{k-j-m}}{\binom{n-i+1}{k-j}}}
= \begin{cases}
    \sqrt{1-m +  \frac{(-1)^m(j-k)}{n-i+1}} & \text{if} \quad k-j \le n-i +1\\
    0 & \text{else}
\end{cases}\,,
\label{gamma}
\end{equation}
and \eqref{DickeMPS} and \eqref{Dickeelem} follows. We can now understand the ansatz \eqref{ansatz}: Eqs. \eqref{step1a} and \eqref{step2} require $j'=j-m$ because $\langle D^{i-1}_{j-m}|D^{i-1}_{j'}\rangle=\delta_{j',j-m}$. 

Strictly speaking, this MPS is not canonical. However, the closely-related matrices $\bar{A}^m_i$ defined by
\begin{align}
    \bar{A}^0_i &= \begin{cases}
       A^0_i & 1 \le i \le n -k+1\\
       A^0_i + \sum_{j=0}^{k+i-n-2} e_{jj} & n-k+2 \le i \le n 
    \end{cases} \,, \nonumber \\
    \bar{A}^1_i &= A^1_i \,, \qquad\qquad \qquad \quad  1 \le i \le n 
    \label{Abar}
\end{align}
{\it do} obey the canonicity condition
$\sum_{m=0}^1 \bar{A}_i^{m\, \dagger}\, \bar{A}^m_i = \id$; moreover, the MPS result \eqref{DickeMPS} is also satisfied if each $A^m_i$ is replaced by $\bar{A}^m_i$.
(The $e_{ab}$ in \eqref{Abar}
are elementary $(k+1) \times (k+1)$ matrices, with matrix elements $\left(e_{ab}\right)_{j_1,j_2} = \delta_{a,j_1}\, \delta_{b,j_2}$.) In other words, the $k-1$ leftmost $A^0_i$ matrices are missing some 1's along the upper part of the diagonal. Nonetheless, our MPS is `sufficiently' canonical, in the sense that there still exist unitaries $U_i$ implementing \eqref{Ugen} for \emph{sufficiently} many values of $j$ in \eqref{Ugen} such that 
\begin{equation}
U_n\dots U_2U_1\,|\underline 0\rangle\,|0\rangle^{\otimes n}=
|\underline k\rangle\,|D^n_k\rangle\,.
\label{sequential}
\end{equation}
This is due to our choice of boundary conditions. For example, since $|\underline R\rangle=|\underline 0\rangle$, it suffices that $U_1$ implements \eqref{Ugen} for just $j=0$; consequently, the only entries in the matrices $A_1^m$ that matter are the ones in the leftmost column, so  setting other columns to zero does not affect the sequential preparation of the state with unitaries. A similar argument can be made that the added matrix elements in $\bar A^{m_i}_i$ have no impact on the sequential preparation of the state, and so our MPS is `sufficiently' canonical. Thus, here and in what follows, we disregard this subtlety, and refer to our MPS representations as canonical.

\subsection{Higher-spin Dicke states}

The qubit Dicke states can be expressed as $|D^n_k\rangle\propto (\mathbb{S}^-)^k\,|0\rangle^{\otimes n}$, where $\mathbb{S}^-$ is the total spin-lowering operator for a system of $n$ spin-1/2 spins. A natural generalization of the qubit Dicke states to higher spin, or spin-$s$ Dicke states, is defined by
\begin{equation}
|D^{(s)}_{n,k}\rangle\propto (\mathbb{S}^-)^k\,|0\rangle^{\otimes n}\,,
\end{equation}
where $\mathbb{S}^-$ is now the total
spin-lowering operator for a system of $n$ spin-$s$ spins,
where $s=1/2, 1, 3/2, \ldots$. 
These multi-qudit states with $d=2s+1$ were studied in detail in \cite{Nepomechie:2024fhm}, and were shown to take the closed form expression
\begin{equation}
|D^{(s)}_{n,k}\rangle= \sum_{\substack{j_i=0,1,\dots,2s\\ j_1+j_2+\dots+j_{n}=k}}
\sqrt{\frac{
\binom{2s}{j_1}
\binom{2s}{j_2}\dots
\binom{2s}{j_{n}}
}{
\binom{2sn}{k}}}\,|j_{n}\dots j_2 j_1\rangle\,.
\label{spinsclosed}
\end{equation}
For example, with $(n,k,s)=(4,2,1)$ we have
\begin{align}
    |D^{(1)}_{4,2}\rangle& = \frac{1}{\sqrt{7}}\left(|00 1 1 \rangle + |0 1 0 1  \rangle +|01 1 0 \rangle+
    |100 1 \rangle + |1010  \rangle +|1 1 00 \rangle
    \right)\\&+
    \frac{1}{2\sqrt{7}}\left(|0002 \rangle + |0020  \rangle +|0200 \rangle+
    |2000 \rangle
    \right)\,.\non
\end{align}
The combinatorial factor in \eqref{spinsclosed} is the multivariate hypergeometric distribution: consider marbles of $n$ colors (sites), and $2s$ marbles (excitations) per color, so that there are $2sn$ total marbles; then the probability of selecting
$j_i$ marbles of the $i$th color for each $i$, given that $k$ marbles were selected from the $2sn$ marbles without replacement, is precisely $\binom{2s}{j_1}
\binom{2s}{j_2}\dots
\binom{2s}{j_{n}}\big/
\binom{2sn}{k}$.

The Schmidt decomposition for the spin-$s$ Dicke states is \cite{Nepomechie:2024fhm}
\begin{equation}
|D^{(s)}_{n,k}\rangle=\sum_{j=\max(0,k-2s(n-i))}^{\min(k,2si)} c^{nks}_{ij}\,|D^{(s)}_{n-i,k-j}\rangle\,|D^{(s)}_{i,j}\rangle\,,
\qquad  c^{nks}_{ij}=\sqrt{\frac{\binom{2si}{j}\binom{2s(n-i)}{k-j}}{\binom{2sn}{k}}}\,.
\label{spinsSchmidt}
\end{equation}
The Schmidt rank is symmetric under $k\to 2sn-k$, and so we restrict $k\leq sn$ for simplicity. Using the same approach as for the spin-1/2 case, one can show that a canonical MPS for the spin-$s$ Dicke states with minimal bond dimension $\chi=k+1$ is 
\begin{equation}
|D^{(s)}_{n,k}\rangle=\sum_{\vec m}\langle \underline{k}|A_n^{m_n}\dots A_2^{m_2}A_1^{m_1}|\underline 0\rangle\,|\vec m\rangle\,,
\label{DickeMPSspins}
\end{equation} 
where $A_i^m$ are $(k+1) \times (k+1)$  matrices
with elements 
\begin{equation}
\langle\underline j'|A_i^{m}|\underline j\rangle=\gamma^{(i)}_{j,m}\,\delta_{j',j+m} \,, \qquad
\gamma^{(i)}_{j,m} = \frac{c^{nks}_{i,j+m}c^{i,j+m,s}_{i-1,j}}{c^{nks}_{i-1,j}} = \sqrt{\frac{\binom{2s}{m} \binom{2s(n-i)}{k-j-m}}{\binom{2s(n-i+1)}{k-j}}}
\,.
\label{Dickeelemspins}
\end{equation}
The unitarity of $U_i$ in \eqref{Ugen} requires that  $\gamma^{(i)}_{j,m}$ should satisfy the constraint
\begin{equation}
\sum_{m=0}^{2s} \left(\gamma^{(i)}_{j,m} \right)^2 =1 \,,
\end{equation}
which, being an example of Vandermonde's identity, is indeed satisfied.

We note the curious fact that the MPS bond dimension $\chi$ of the Dicke state $|D^{(s)}_{n,k}\rangle$ depends only on the value of $k$; in particular, $\chi$ does not depend on the value of $s$, even though the local Hilbert space dimension grows as $s$. A heuristic understanding of this fact can be gleaned from a simpler (translationally invariant, non-canonical) MPS, namely,
\begin{equation}
    A^m = c_m\, (S^-)^m \,, 
    \qquad   c_m = \sqrt{\frac{\binom{2s}{m}}{(2s)^m}} \,,
    \qquad m = 0, \ldots 2s \,,
    \label{simpler}
\end{equation}
where $S^-$ is the $(k+1) \times (k+1)$ spin-lowering 
operator corresponding to spin $s' \equiv k/2$.
Indeed, since $S^- |\underline{j}\rangle \propto |\underline{j+1}\rangle$, 
it is clear that the matrix element 
$\langle \underline{k}|A^{m_n}\dots A^{m_2}A^{m_1}|\underline 0\rangle$
in \eqref{DickeMPSspins} vanishes unless $\sum_i m_i = k$, in accordance with \eqref{spinsclosed}. We see that $\chi$ (the size of the $A$-matrices in \eqref{simpler})
depends on $s'$, rather than on $s$.

\section{$SU(d)$ Dicke states}\label{sec:sud}

Similar to $SU(2)$ Dicke states where the number of 1's is fixed, in $SU(d)$ Dicke states, the occupation number
for each level (i.e., how many qudits are in each of the $d$ levels) is fixed. Given a
specification of occupation numbers for each level, a qudit Dicke state is a uniform superposition of
all qudit computational basis states that achieve this distribution of occupation numbers.

More explicitly,
let $\vec k = (k_{0}, k_{1}, \ldots, k_{d-1})$ be a $d$-dimensional vector whose components are integers from 0 to $n$ (that is, $k_i \in \{0, 1, \ldots, n \}$) that sum to $n$ (that is,  $\sum_{j=0}^{d-1} k_{j} = n$).
We consider the corresponding qudit Dicke state $|D^{n}(\vec k)\rangle$ of $n$ $d$-level qudits \cite{Wei:2003, Popkov:2005, Hayashi:2008,  Wei:2008, Zhu:2010, Carrasco:2015sxh,Aloy:2021, Li:2021, Nepomechie:2023lge}
\begin{equation}
|D^{n}(\vec k)\rangle  = \frac{1}{\sqrt{{n \choose \vec k}}}
\sum_{w \in \mathfrak{S}_{M(\vec k)}} | w \rangle \,,
\label{quditDickedef}
\end{equation}
where $\mathfrak{S}_{M(\vec k)}$ is the set of permutations of the 
multiset $M(\vec k)$ 
\begin{equation}
M(\vec k)	=\{ \underbrace{0, \ldots, 0}_{k_{0}}, \underbrace{1, 
\ldots, 1}_{k_{1}}, \ldots, \underbrace{d-1, \ldots, d-1}_{k_{d-1}}\} 
\,,
\label{multiset}
\end{equation}
where $k_{j}$ is the multiplicity of $j$ in $M(\vec k)$, such that $M(\vec 
k)$ has cardinality $n$;
and $| w  \rangle$ is the state of $n$ qudits corresponding 
to the permutation $w$. Moreover, ${n \choose \vec k}$ denotes the multinomial
\begin{equation}
{n \choose \vec k} = {n \choose k_{0}, k_{1}, \ldots, k_{d-1}}	= 
\frac{n!}{\prod_{j=0}^{d-1}k_{j}!} \,,
\label{size}
\end{equation}
which is the cardinality of $\mathfrak{S}_{M(\vec k)}$. An example with $\vec k = (2,1,1)$, so that $d=3$ (qutrits) and $n=4$, is
\begin{align}
|D^{4}(2,1,1)\rangle  &= \frac{1}{\sqrt{12}}
\Big(|0012\rangle + |0102\rangle +
|1002\rangle + 
|0021\rangle + |0201\rangle + |2001\rangle \non \\
&+ |0210\rangle + |0120\rangle + |1020\rangle + 
|1200\rangle + |2010\rangle + |2100\rangle \Big)\,. 
\label{d3example}
\end{align}

The qudit Dicke state $|D^{n}(\vec k)\rangle$ has the following Schmidt decomposition in terms of qudit Dicke states of size $n-l$ and $l$ \cite{Carrasco:2015sxh, Raveh:2023iyy}
\begin{equation}
|D^{n}(\vec k)\rangle  = \sum_{\vec{a} \in \mathcal{A}^l(\vec{k})} 
c^{n,\vec k}_{l, \vec{a}}\, |D^{n-l}(\vec{k}-\vec{a})\rangle \,|D^{l}(\vec{a})\rangle \,, \qquad c^{n,\vec k}_{l, \vec{a}} = \sqrt{\frac{\binom{l}{\vec a} \binom{n-l}{\vec{k}-\vec{a}}}{\binom{n}{\vec{k}}}} \,, \qquad l = 1, \ldots, n\,,
\end{equation}
where the set $\mathcal{A}^l(\vec{k})$ is defined by
\begin{equation}
    \mathcal{A}^l(\vec{k}) = \big\{ \vec{a} = (a_0, a_1, \ldots, a_{d-1})\, \big\vert\quad 0 \le a_i \le k_i \,, \quad \sum_{i=0}^{d-1} a_i = l \big\} \,,
    \label{setA}
\end{equation}
whose cardinality we denote by $\mathcal{D}^l(\vec{k}) = \big\vert  \mathcal{A}^l(\vec{k})  \big\vert$.\footnote{The cardinality of the set $\mathcal{A}^l(\vec k)$ is known to be difficult to express in closed form, see e.g. \cite{MSEextended}.} As $l$ varies from 1 to $n$, $\mathcal{D}^l(\vec{k})$ attains its maximum value for $l=\lfloor n/2 \rfloor$, so the minimum bond dimension of an MPS for $|D^{n}(\vec k)\rangle$ is  $\mathcal{D}^{\lfloor n/2 \rfloor}(\vec{k})$. We now proceed to construct a canonical MPS with this bond dimension. Examples of values of $\chi = \mathcal{D}^{\lfloor n/2 \rfloor}(\vec{k})$ for small values of $n$ and $d=3$ are listed in Table \ref{table:dims}.

\begin{table}[htb]
\centering
\begin{tabular}{|c|c|c|}
$\vec{k}$ & $n$ & $\chi$\\
\hline
$(1,1,1)$ & 3 & 3 \\
$(1,1,2)$ & 4 & 4 \\
$(1,1,3)$ & 5 & 4 \\
$(1,2,2)$ & 5 & 5 \\
$(2,2,2)$ & 6 & 7 \\
$(1,3,3)$ & 7 & 7 \\
$(2,3,3)$ & 8 & 10
\end{tabular}
\caption{Examples of values of $\chi=\mathcal{D}^{\lfloor n/2 \rfloor}(\vec{k})$ for small values of $n$ and $d=3$.}
\label{table:dims}
\end{table}

In the $SU(2)$ case, each Dicke state is naturally associated with an integer corresponding to a basis state of the ancilla: $|D^i_j\rangle\leftrightarrow |\underline j\rangle$, see \eqref{step1}. For qudit Dicke states, these is no natural way to similarly associate each $|D^l(\vec a)\rangle$ with an integer; instead, we simply assign integers to index the various possible $\vec a$'s. Let the elements of $\mathcal{A}^l(\vec{k})$ \eqref{setA} be labeled by consecutive integers $j^l(\vec{a}) = 0, 1, \ldots, \mathcal{D}^l(\vec{k})-1$ in an arbitrary manner.
For $\vec{a} \in \mathcal{A}^{l-1}(\vec{k})$, we make an ansatz generalizing \eqref{ansatz}
\begin{equation}
U_l\, |\underline{j^{l-1}(\vec{a})}\rangle\, |0\rangle_l 
= \sum_{m=0}^{d-1} \left( A^m_l\, |\underline{j^{l-1}(\vec{a})}\rangle \right)
|m\rangle_l 
= \sum_{m=0}^{d-1} \gamma^{(l)}_{j^{l-1}(\vec{a}),m}\,  
|\underline{j^{l}(\vec{a} + \hat{m})}\rangle\, |m\rangle_l  \,,
\end{equation}
where $\hat m$ is a $d$-dimensional unit vector that has components $(\hat m)_i =\delta_{m,i}$, with $m=0, 1, \dots, d-1$.
Note that $\vec{a} + \hat{m} \in \mathcal{A}^{l}(\vec{k})$ if and only if $a_m<k_m$. If  $\vec{a} + \hat{m} \notin \mathcal{A}^{l}(\vec{k})$, then $\gamma^{(l)}_{j^{l-1}(\vec{a}),m} =0$. 
The corresponding MPS 
with bond dimension $\chi=\mathcal{D}^{\lfloor n/2 \rfloor}(\vec{k})$ is 
\begin{equation}
|D^n(\vec{k})\rangle=\sum_{\vec m}\langle \underline{0}|A_n^{m_n}\dots A_2^{m_2}A_1^{m_1}|\underline 0\rangle\,|\vec m\rangle\,,
\label{quditDickeMPSspins}
\end{equation} 
where $A_l^m$ are $\chi \times \chi$  matrices
with elements 
\begin{equation}
    \langle \underline{j^l(\vec{a'})}| A_l^m | \underline{j^{l-1}(\vec{a})} \rangle = \gamma^{(l)}_{j^{l-1}(\vec{a}),m}\, \delta_{\vec{a'},\vec{a}+\hat{m}} \,,
    \label{Amatelemsqudit}
\end{equation}
where $\vec{a} \in \mathcal{A}^{l-1}(\vec{k})$ and $\vec{a'} \in \mathcal{A}^{l}(\vec{k})$. Proceeding as before, we obtain
\begin{equation}
\gamma^{(l)}_{j^{l-1}(\vec{a}),m} =  \frac{c^{n,\vec k}_{l,\vec{a}+\hat m}c^{l,\vec a+\hat m}_{l-1,\vec{a}}}{c^{n,\vec k}_{l-1,\vec{a}}}
 \,.
\label{gammaqudit}
\end{equation}

\section{Discussion}\label{sec:discuss}

Our main results are MPS representations for qubit, higher-spin, and qudit Dicke states, see Eqs. \eqref{Dickeelem}, \eqref{gamma}, \eqref{Dickeelemspins}, \eqref{Amatelemsqudit}, \eqref{gammaqudit}.
We emphasize that these MPS representations are  exact, canonical, and have minimal bond dimension. (These representations are `sufficiently' canonical in the sense that they can be used for sequential state preparation; they can be made strictly canonical by adding suitable correction terms, as in Eq. \eqref{Abar} for the qubit case.)

Dicke states are translational invariant (TI) states, since they are invariant under the one-site shift operator. However, the MPS formulations obtained here are not TI.\footnote{It is often the case that a mathematical treatment of a problem does not respect all of its symmetries. A well-known example is the quantization of a gauge-invariant theory: it is convenient to specify a gauge, which apparently breaks gauge invariance; nevertheless, physical quantities remain gauge-invariant.} Taking   \eqref{simpler} instead of \eqref{Dickeelem} gives a TI MPS with open boundary conditions (OBC) and bond dimension $\chi=k+1$, which however is {\it not} canonical. A canonical TI MPS with periodic boundary conditions (PBC) {\it can} be obtained via the construction in the proof of Theorem 3 in \cite{Perez-Garcia:2006nqo}, at the expense of increasing the bond dimension from $\chi$ to $n\chi$. It is conjectured that there does \emph{not} exist a $\chi=O(1)$ TI MPS with PBC for the W-state, which is the simplest Dicke state $|D^n_k\rangle$ with $k=1$, see Appendix A.1 in \cite{Perez-Garcia:2006nqo} and also \cite{Klimov:2023srk}. The apparent clash between translational invariance and canonicity for a given bond dimension may merit further investigation.

We note that the MPS formulations obtained here could be further generalized to so-called $q$-analogs of Dicke states: namely, generalizations of Dicke states introduced in \cite{Li:2015, Raveh:2023iyy} that involve a (complex) parameter $q$, which reduce to usual Dicke states in the limit $q \rightarrow 1$. These $q$-analog Dicke states include other studied states as special cases.
For example, antisymmetric states \cite{Wei:2003, Hayashi:2008, Li:2021} correspond to the particular case $q=-1$. The Schmidt decompositions for the $SU(2)$ and $SU(d)$ cases, which are necessary for deriving the corresponding MPS formulations, were already found in \cite{Li:2015} and \cite{Raveh:2023iyy}, respectively.

As described in Section \ref{sec:general} and Appendix \ref{sec:circuit}, the canonical MPS for the Dicke states naturally leads to a sequential algorithm for their deterministic preparation on a quantum computer, with $O(k\, n)$ size and depth.
This algorithm is closely related to the algorithm in \cite{Bartschi2019}, where the recursive nature of the Dicke states is leveraged to formulate a quantum algorithm for their preparation. The primary difference between the two circuits is the manner in which the rotation operations are controlled; in  the circuit in Appendix \ref{sec:circuit}, the controls are implemented via the qudit ancilla, while the circuit in \cite{Bartschi2019} manages to avoid any ancillas. However, the values of the rotation angles in the two circuits coincide exactly; in this sense, we have brought the algorithm in \cite{Bartschi2019} into the MPS framework. 

As noted in the Introduction, a constant-depth quantum circuit for preparing AKLT states deterministically has recently been formulated using the exact MPS together with intermediate measurement and feed-forward techniques \cite{Smith:2023}, and has been generalized to MPS states with global on-site symmetries \cite{Smith:2024}. It would be very interesting to similarly formulate a constant-depth quantum circuit for preparing Dicke states deterministically using their exact MPS. Exploiting quantum phase estimation as in \cite{Wang:2021}, a 
probabilistic algorithm for exactly preparing Dicke states $|D^n_k\rangle$ in $O(\log_2 n)$ depth with success probability $O(k^{-1/2})$ has recently been proposed in \cite{Piroli:2024ckr}.

\section*{Acknowledgments} 

RN is supported in part by the National Science Foundation under grant PHY 2310594, and by a Cooper fellowship. 

\appendix

\section{Sequential preparation of Dicke states}\label{sec:circuit}

We present here an explicit quantum circuit for the deterministic sequential preparation of the qubit Dicke state $|D^n_k\rangle$ \eqref{Dickedef} based on the MPS \eqref{genprep}, \eqref{DickeMPS}. The key step is to formulate a quantum circuit for the unitary operator $U_i$ in Eqs. \eqref{ansatz}, \eqref{gamma}. Inspired by \cite{Bartschi2019} (see also \cite{Nepomechie:2023lge, Nepomechie:2024fhm}), we assume that $U_i$ can be expressed as an ordered product of simpler operators $I^{(i)}_l$
\begin{equation}
    U_i =  \overset{\curvearrowleft}{\prod_{l=0}^{k}} I^{(i)}_l \,,
    \label{UI}
\end{equation}
where the product goes from right to left with increasing $l$, and the operator $I^{(i)}_l$ performs the mapping \eqref{ansatz} on $|\underline j\rangle\,|0\rangle_i$ if $l=j$
\begin{equation}
I^{(i)}_l\,|\underline j\rangle\,|0\rangle_i 
= \begin{cases}
   |\underline j\rangle\,|0\rangle_i  & l\ne j \\[0.1in]
   \sum_{m=0}^1\gamma^{(i)}_{j,m}\,|\underline{j+m}\rangle\,|m\rangle_i & l=j
\end{cases} \,;
\label{Il}
\end{equation}
moreover
\begin{equation}
I^{(i)}_l\,|\underline j\rangle\,|1\rangle_i =
|\underline j\rangle\,|1\rangle_i \,, \qquad\qquad l \ne j-1 \,.
\end{equation}
It follows that the $I^{(i)}_l$ operators do not interfere with each other:
\begin{equation}
    I^{(i)}_l \left(I^{(i)}_j |\underline j\rangle\,|0\rangle_i \right) = \left(I^{(i)}_j |\underline j\rangle\,|0\rangle_i \right)  \qquad l>j \,.
    \label{nointerfere}
\end{equation}
We shall see that the product over $l$ in \eqref{UI} can be restricted, see \eqref{UIsimple} below.

\begin{figure}[htb]
	\centering
\begin{adjustbox}{width=0.3\textwidth}
\begin{quantikz}
\lstick{$|0\rangle_i$} & \gateO{0} \vqw{1} & \gate{R(\theta)} \vqw{1} 
& \gateO{0} \vqw{1} & \qw \\
\lstick{$|\underline{j} \rangle$} & \gate{\oplus}  & \gateO{l+1} & \gate{\ominus}  & \qw \\
\end{quantikz}
\end{adjustbox}
\caption{Circuit diagram for $I^{(i)}_l \,|\underline j\rangle\,|0\rangle_i $}
\label{fig:I}
\end{figure}
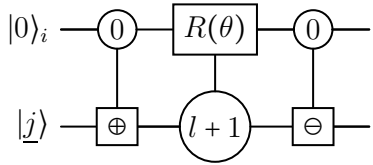

The operator $I^{(i)}_l$ can be implemented by the quantum circuit whose  circuit diagram is shown in Fig. \ref{fig:I}.
In this figure, the top wire represents qubit $i$, while the bottom wire represents the $(k+1)$-level qudit ancilla. The circles $\begin{quantikz}\gateO{i}\end{quantikz}$ denote controls. The 1-qubit rotation gate $R(\theta)$, which is given by
\begin{equation}
R(\theta)  
 = \begin{pmatrix}
       \cos(\theta/2) & -\sin(\theta/2) \\[0.1 cm]
       \sin(\theta/2) &  \cos(\theta/2)
\end{pmatrix} \,, \qquad \cos(\theta/2) = \gamma^{(i)}_{j,0} 
 \,, \qquad \sin(\theta/2) = \gamma^{(i)}_{j,1} \,,
\label{Rgate}
\end{equation}
is controlled by the ancilla, where the control value $l+1$ is understood as ${\rm mod}(k+1)$. We remind that $\gamma^{(i)}_{j,m}$ is given in \eqref{gamma}.
The one-qudit gates $\oplus$ and $\ominus$, which are defined by
\begin{equation}
    \oplus\, |\underline{j} \rangle = |\underline{j+1} \rangle \,, \qquad
    \ominus\, |\underline{j} \rangle = |\underline{j-1} \rangle \,, \qquad j = 0, 1, \ldots, k\,,
\end{equation}
are controlled by the system qubit $i$, where $j \pm 1$ is understood as ${\rm mod}(k+1)$. (Alternatively, instead of $\oplus$ and $\ominus$, one can use NOT gates $X^{(l,l+1)}$ that map $|\underline{l} \rangle \leftrightarrow |\underline{l+1} \rangle$, and leave invariant other basis states $|\underline{j} \rangle$ with $j\ne l, l+1$.)
Indeed, it is straightforward to check that this circuit satisfies both properties \eqref{Il} and \eqref{nointerfere}. 

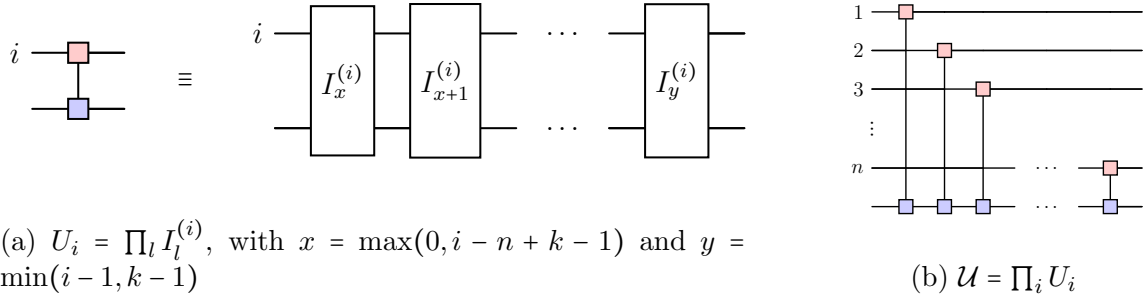
\begin{figure}[htb]
	\centering
	\begin{subfigure}{0.60\textwidth}
      \centering
\begin{adjustbox}{width=1.0\textwidth, raise=2em}
\begin{quantikz}
\lstick{$i$} & \gate[style={fill=red!20}]{} \vqw{1} & \qw \rstick[2, brackets=none]{$\quad\equiv\quad$}\\
\lstick{} & \gate[style={fill=blue!20}]{}  & \qw \\
\end{quantikz}
\begin{quantikz}
\lstick{$i$} & \gate[2]{I^{(i)}_x} & \gate[2]{I^{(i)}_{x+1}} & \quad\ldots\quad
& \gate[2]{I^{(i)}_y} & \qw \\
\lstick{{}} & \qw  & \qw & \quad\ldots\quad & \qw  & \qw \\
\end{quantikz}
\end{adjustbox}
\caption{}
\label{fig:Ui}
    \end{subfigure}%
    \begin{subfigure}{0.4\textwidth}
        \centering
\begin{adjustbox}{width=0.6\textwidth}
\begin{quantikz}
\lstick{$1$} & \gate[style={fill=red!20}]{} \vqw{5} & \qw & \qw & \qw & \qw & \qw\\
\lstick{$2$} & \qw & \gate[style={fill=red!20}]{} \vqw{4} & \qw & \qw & \qw & \qw\\
\lstick{$3$} & \qw & \qw & \gate[style={fill=red!20}]{} \vqw{3} & \qw & \qw & \qw\\
\vdots \\
\lstick{$n$} & \qw & \qw & \qw &\quad \ldots\quad & \gate[style={fill=red!20}]{} \vqw{1} & \qw \\
\lstick{{}} & \gate[style={fill=blue!20}]{}  & \gate[style={fill=blue!20}]{} & \gate[style={fill=blue!20}]{}&\quad \ldots\quad & \gate[style={fill=blue!20}]{} & \qw \\
\end{quantikz}
\end{adjustbox}
\caption{}
\label{fig:calU}
	 \end{subfigure}	
\caption{Circuit diagram for sequential Dicke state preparation. (a) $U_i=\prod_l I^{(i)}_l$, with $x=\max(0,i-n+k-1)$ and $y=\min(i-1,k-1)$; (b)  $\mathcal{U}=\prod_i U_i$ }
\label{fig:Dicke}
\end{figure}

The unitary operator that prepares the Dicke state $|D^n_k\rangle$ deterministically from the initial state $|\underline 0\rangle\,|0\rangle^{\otimes n}$
is given by an ordered product of the $U_i$ operators
\begin{equation}
    \mathcal{U} = \overset{\curvearrowleft}{\prod_{i=1}^{n}} U_i \,,
    \label{calU}
\end{equation}
since $\mathcal{U}\,|\underline 0\rangle\,|0\rangle^{\otimes n} =  |\underline k\rangle\, |D^n_k\rangle$, see \eqref{sequential}. It is possible to show that the product over $l$ in \eqref{UI} can be restricted, so that $U_i$ is given by
\begin{equation}
U_i =  \overset{\curvearrowleft}{\prod_{l=\max(0,i-n+k-1)}^{\min(i-1,k-1)}} I^{(i)}_l \,,
\label{UIsimple}
\end{equation}
Let us represent $U_i$ \eqref{UIsimple} by the circuit diagram in Fig. \ref{fig:Ui}, where the top wire represents qubit $i$, and the bottom wire represents the $(k+1)$-level qudit ancilla. The operator $\mathcal{U}$ \eqref{calU} can then be represented by the circuit diagram in Fig. \ref{fig:calU}. The size and depth of the circuit is $3k(n+1-k)$, as follows from \eqref{calU} and \eqref{UIsimple}. Code in cirq \cite{cirq} for simulating this circuit is provided in the Supplemental Material \cite{SuppMat}.

\clearpage

\providecommand{\href}[2]{#2}\begingroup\raggedright\endgroup

\end{document}